\documentclass[a4paper,amsfonts,amssymb,amsmath, reprint,showkeys,superscriptaddress,nofootinbib, twoside,longbibliography]{revtex4-2}
\usepackage[english]{babel}
\usepackage[utf8]{inputenc}
\usepackage[colorinlistoftodos, color=green!40, prependcaption]{todonotes}
\usepackage{amsthm}
\usepackage{mathtools}
\usepackage{physics}
\usepackage{xcolor}
\usepackage{graphicx}
\usepackage{adjustbox}
\usepackage{placeins}
\usepackage[T1]{fontenc}
\usepackage{lipsum}
\usepackage{csquotes}

\usepackage[subrefformat=parens,labelformat=parens,caption=false]{subfig}
\usepackage{collref}

\usepackage[pdftex, pdftitle={Article}, pdfauthor={Author},
            colorlinks = true,
            linkcolor = blue,
            urlcolor  = blue,
            citecolor = blue,
            anchorcolor = blue]{hyperref}

\begin{document}

\title{Exceptional-point-assisted entanglement, squeezing, and reset in a chain of three superconducting resonators}

\author{Wallace S. Teixeira}
    \affiliation{Department of Applied Physics, QCD Laboratories, QTF Centre of Excellence, Aalto University, P.O. Box 15100, FI-00076 Aalto, Finland}

\author{Vasilii Vadimov}
    \affiliation{Department of Applied Physics, QCD Laboratories, QTF Centre of Excellence, Aalto University, P.O. Box 15100, FI-00076 Aalto, Finland}

\author{Timm M\"orstedt}
    \affiliation{Department of Applied Physics, QCD Laboratories, QTF Centre of Excellence, Aalto University, P.O. Box 15100, FI-00076 Aalto, Finland}
    
\author{Suman Kundu}
    \affiliation{Department of Applied Physics, QCD Laboratories, QTF Centre of Excellence, Aalto University, P.O. Box 15100, FI-00076 Aalto, Finland}    
    
\author{Mikko M\"ott\"onen}
    \affiliation{Department of Applied Physics, QCD Laboratories, QTF Centre of Excellence, Aalto University, P.O. Box 15100, FI-00076 Aalto, Finland}
    \affiliation{VTT Technical Research Centre of Finland Ltd., QTF Center of Excellence, P.O. Box 1000, FI-02044 VTT, Finland}

\begin{abstract}
The interplay between coherent and dissipative dynamics required in various control protocols of quantum technology has motivated studies of open-system degeneracies, referred to as exceptional points (EPs). Here, we introduce a scheme for fast quantum-state synthesis using exceptional-point engineering in a lossy chain of three superconducting resonators. We theoretically find that the rich physics of EPs can be used to identify regions in the parameter space that favor a fast and quasi-stable transfer of squeezing and entanglement or a fast reset of the system. For weakly interacting resonators with the coupling strength $g$, the obtained quasi-stabilization time scales are identified as $1/(2\sqrt{2}g)$, and reset infidelities below $10^{-5}$ are obtained with a waiting time of roughly $6/g$ in the case of weakly squeezed resonators. Our results shed light on the role of EPs in multimode Gaussian systems and pave the way for optimized distribution of squeezing and entanglement between different nodes of a photonic network using dissipation as a resource.

\end{abstract}


\global\long\def\ket#1{|#1\rangle}%

\global\long\def\Ket#1{\left|#1\right>}%

\global\long\def\bra#1{\langle#1|}%

\global\long\def\Bra#1{\left<#1\right|}%

\global\long\def\bk#1#2{\langle#1|#2\rangle}%

\global\long\def\BK#1#2{\left\langle #1\middle|#2\right\rangle }%

\global\long\def\kb#1#2{\ket{#1}\!\bra{#2}}%

\global\long\def\KB#1#2{\Ket{#1}\!\Bra{#2}}%

\global\long\def\mel#1#2#3{\bra{#1}#2\ket{#3}}%

\global\long\def\MEL#1#2#3{\Bra{#1}#2\Ket{#3}}%

\global\long\def\n#1{|#1|}%

\global\long\def\N#1{\left|#1\right|}%

\global\long\def\ns#1{|#1|^{2}}%

\global\long\def\NS#1{\left|#1\right|^{2}}%

\global\long\def\nn#1{\lVert#1\rVert}%

\global\long\def\NN#1{\left\lVert #1\right\rVert }%

\global\long\def\nns#1{\lVert#1\rVert^{2}}%

\global\long\def\NNS#1{\left\lVert #1\right\rVert ^{2}}%

\global\long\def\ev#1{\langle#1\rangle}%

\global\long\def\EV#1{\left\langle #1\right\rangle }%

\global\long\def\ha{\hat{a}}%

\global\long\def\hb{\hat{b}}%

\global\long\def\hc{\hat{c}}%

\global\long\def\hd{\hat{d}}%

\global\long\def\he{\hat{e}}%

\global\long\def\hf{\hat{f}}%

\global\long\def\hg{\hat{g}}%

\global\long\def\hh{\hat{h}}%

\global\long\def\hi{\hat{i}}%

\global\long\def\hj{\hat{j}}%

\global\long\def\hk{\hat{k}}%

\global\long\def\hl{\hat{l}}%

\global\long\def\hm{\hat{m}}%

\global\long\def\hn{\hat{n}}%

\global\long\def\ho{\hat{o}}%

\global\long\def\hp{\hat{p}}%

\global\long\def\hq{\hat{q}}%

\global\long\def\hr{\hat{r}}%

\global\long\def\hs{\hat{s}}%

\global\long\def\hu{\hat{u}}%

\global\long\def\hv{\hat{v}}%

\global\long\def\hw{\hat{w}}%

\global\long\def\hx{\hat{x}}%

\global\long\def\hy{\hat{y}}%

\global\long\def\hz{\hat{z}}%

\global\long\def\hA{\hat{A}}%

\global\long\def\hB{\hat{B}}%

\global\long\def\hC{\hat{C}}%

\global\long\def\hD{\hat{D}}%

\global\long\def\hE{\hat{E}}%

\global\long\def\hF{\hat{F}}%

\global\long\def\hG{\hat{G}}%

\global\long\def\hH{\hat{H}}%

\global\long\def\hI{\hat{I}}%

\global\long\def\hJ{\hat{J}}%

\global\long\def\hK{\hat{K}}%

\global\long\def\hL{\hat{L}}%

\global\long\def\hM{\hat{M}}%

\global\long\def\hN{\hat{N}}%

\global\long\def\hO{\hat{O}}%

\global\long\def\hP{\hat{P}}%

\global\long\def\hQ{\hat{Q}}%

\global\long\def\hR{\hat{R}}%

\global\long\def\hS{\hat{S}}%

\global\long\def\hT{\hat{T}}%

\global\long\def\hU{\hat{U}}%

\global\long\def\hV{\hat{V}}%

\global\long\def\hW{\hat{W}}%

\global\long\def\hX{\hat{X}}%

\global\long\def\hY{\hat{Y}}%

\global\long\def\hZ{\hat{Z}}%

\global\long\def\hap{\hat{\alpha}}%

\global\long\def\hbt{\hat{\beta}}%

\global\long\def\hgm{\hat{\gamma}}%

\global\long\def\hGm{\hat{\Gamma}}%

\global\long\def\hdt{\hat{\delta}}%

\global\long\def\hDt{\hat{\Delta}}%

\global\long\def\hep{\hat{\epsilon}}%

\global\long\def\hvep{\hat{\varepsilon}}%

\global\long\def\hzt{\hat{\zeta}}%

\global\long\def\het{\hat{\eta}}%

\global\long\def\hth{\hat{\theta}}%

\global\long\def\hvth{\hat{\vartheta}}%

\global\long\def\hTh{\hat{\Theta}}%

\global\long\def\hio{\hat{\iota}}%

\global\long\def\hkp{\hat{\kappa}}%

\global\long\def\hld{\hat{\lambda}}%

\global\long\def\hLd{\hat{\Lambda}}%

\global\long\def\hmu{\hat{\mu}}%

\global\long\def\hnu{\hat{\nu}}%

\global\long\def\hxi{\hat{\xi}}%

\global\long\def\hXi{\hat{\Xi}}%

\global\long\def\hpi{\hat{\pi}}%

\global\long\def\hPi{\hat{\Pi}}%

\global\long\def\hrh{\hat{\rho}}%

\global\long\def\hvrh{\hat{\varrho}}%

\global\long\def\hsg{\hat{\sigma}}%

\global\long\def\hSg{\hat{\Sigma}}%

\global\long\def\hta{\hat{\tau}}%

\global\long\def\hup{\hat{\upsilon}}%

\global\long\def\hUp{\hat{\Upsilon}}%

\global\long\def\hph{\hat{\phi}}%

\global\long\def\hvph{\hat{\varphi}}%

\global\long\def\hPh{\hat{\Phi}}%

\global\long\def\hch{\hat{\chi}}%

\global\long\def\hps{\hat{\psi}}%

\global\long\def\hPs{\hat{\Psi}}%

\global\long\def\hom{\hat{\omega}}%

\global\long\def\hOm{\hat{\Omega}}%

\global\long\def\hdgg#1{\hat{#1}^{\dagger}}%

\global\long\def\cjg#1{#1^{*}}%

\global\long\def\hsgx{\hat{\sigma}_{x}}%

\global\long\def\hsgy{\hat{\sigma}_{y}}%

\global\long\def\hsgz{\hat{\sigma}_{z}}%

\global\long\def\hsgp{\hat{\sigma}_{+}}%

\global\long\def\hsgm{\hat{\sigma}_{-}}%

\global\long\def\hsgpm{\hat{\sigma}_{\pm}}%

\global\long\def\hsgmp{\hat{\sigma}_{\mp}}%

\global\long\def\dert#1{\frac{d}{dt}#1}%

\global\long\def\dertt#1{\frac{d#1}{dt}}%

\global\long\def\Tr{\text{Tr}}%

\maketitle

\section{Introduction}\label{sec:intro}

Quantum mechanics has provided profoundly novel ways of information processing, communication, and metrology~\cite{Dowling2003}. Although nonlinearity expressed by the anharmonicity of energy levels is a key metric for physical realizations of qubits, quantum harmonic systems have also a broad range of quantum-technological applications employing, e.g., squeezing and entanglement as resources~\cite{Weedbrook2012,Serafini2017}. 
The efficient use of such properties in experiments typically requires quick transitions from coherent to incoherent dynamics for different stages of the protocols, and hence dissipation engineering using \textit{in-situ} tunable components plays an important role towards fast control and scalability of practical quantum systems~\cite{Chen2018}.

In circuit quantum electrodynamics (cQED), for example, efforts have been made to integrate devices with \textit{in-situ}-tunable dissipation to prepare specific quantum states~\cite{Murch2012,Shankar2013,Holland2015,Leghtas2015,KimchiSchwartz2016,Premaratne2017,Lu2017,Dassonneville2021}, produce fast reset~\cite{Valenzuela2006,Geerlings2013,Tan2017,Silveri2017,Magnard2018,Partanen2018,Sevriuk2019,Yoshioka2021,Zhou2021,Vadimov2022,Moerstedt2022,Sevriuk2022}, and to exploit the potential benefits of open-system degeneracies, referred to as exceptional points (EPs)~\cite{Magnard2018,Partanen2019,Naghiloo2019,Zhou2021,Chen2021,Chen2022,Abbasi2022}. In contrast to Hermitian degeneracies, EPs induce the coalescence of eigenvalues and eigenvectors of the dynamical matrix governing the open-system evolution leading to critical dynamics manifested by polynomial solutions in time~\cite{AmShallem2015,Minganti2019}. These features are key elements for optimized heat flow~\cite{Partanen2019} and sensitive parameter estimation~\cite{AmShallem2015}. When EPs are dynamically encircled in the parameter space, counterintuitive effects not observed in closed systems appear, such as the breakdown of the adiabatic approximation and topological energy transfer~\cite{Uzdin2011,Milburn2015,Xu2016}. Due to their novelty for the observation of open-system phenomena and applications, EPs have also been acknowledged in other physical architectures~\cite{Ding2021,Miri2019,Hodaei2017}. 
However, the relationship between EPs and the emergence of nonclassical and nonlocal features in multipartite continuous-variable (CV) quantum systems has not been fully explored~\cite{VashahriGhamsari2017,Chakraborty2019,Perina2019,Kalaga2019,Roccati2021,Roy2021}.

Quantum harmonic arrays have a practical appeal in cQED for the implementation of quantum memories~\cite{Naik2017} and for the capability to simulate many-body physics~\cite{Underwood2012}. Even though the transport of quantum correlations has been extensively theoretically studied in related setups~\cite{Audenaert2002,Plenio2005,Leandro2009,Nicacio2016}, the high dimension of such systems and their dissipative features render the characterization of EPs an involved procedure~\cite{Ryu2012,Wu2018,Downing2021}.

Motivated by the above-mentioned potential use cases and issues, in this paper, we introduce exceptional-point engineering for squeezing and entanglement propagation. 
We consider a minimal setup for the production of high-order EPs, consisting of a chain of three linearly coupled superconducting resonators with independent decay channels.
To some extent, our system can be described by its first and second moments, so that it can constitute an example of a Gaussian system, i.e., a CV system represented by a Gaussian Wigner function~\cite{Serafini2017}. To analytically describe the EP-related phenomena, we employ the Jordan normal form of the dynamical matrix of the second moments, allowing for investigations beyond energy flow. 

Interestingly, we observe that even for weakly coupled resonators, the operation in the vicinity of a specific second-order EP may turn the central resonator into a fast squeezing splitter and distant-entanglement generator using only initial squeezing in a single resonator. We calculate theoretical bounds for the squeezing and entanglement of the quasistable states and observe their rich dependence on the initial squeezing parameter. The entanglement generation here relies on the availability of initial squeezing since the beam-splitter-type interactions do not entangle the resonators in coherent states. On the other hand, operation near a different, third-order EP branch provides substantial speed up of the decay towards the ground state. Therefore, the detailed knowledge of its open-system degeneracies renders the system a versatile structure for protocols requiring fast stabilization or reset of the desired properties. 

This article is organized as follows. In Sec.~\ref{sec:modela}, we present the general theory of exceptional points in noisy Gaussian systems. In Sec.~\ref{sec:modelb}, we provide the details of the considered setup, including the characterization of its EPs. Sections~\ref{sec:resultsa} and~\ref{sec:resultsb} are dedicated to the studies of different effects arising at or near EPs with a focus on the quasistabilization and decay of nonclassical Gaussian states, respectively. A discussion on the use cases and limitations of EP engineering is provided in Sec.~\ref{sec:disc}. The conclusions are drawn in Sec.~\ref{sec:conc}.

\section{Exceptional points in noisy Gaussian systems}\label{sec:modela}

\begin{figure}
    \subfloat{\label{fig:modela}} 
	\subfloat{\label{fig:modelb}}
    \centering
    \includegraphics[width=\linewidth]{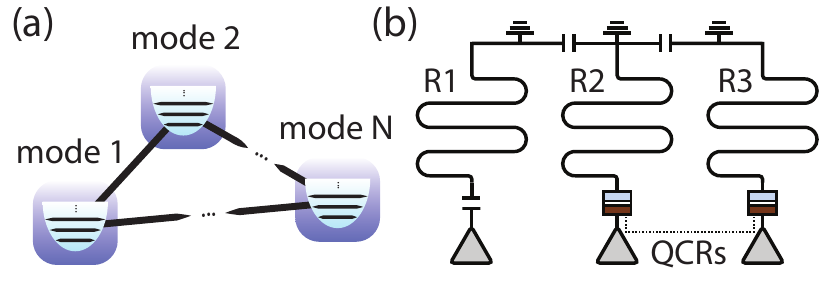}
    \caption{(a) Schematic of the general system considered in this paper consisting of $N$- harmonic quantum modes linearly coupled to one another (black lines). In addition, each mode is coupled to their own Markovian environment (rounded squares). (b) Particular realization of the system explored in this paper, where three superconducting resonators are capacitively coupled in a linear-chain configuration. In addition, each resonator has their own drive lines (triangles), using which the system can be prepared and measured. The decay rates of resonators $R2$ and $R3$ can be controlled by quantum-circuit refrigerators (QCRs) placed at the resonator input ports. Each QCR is composed of a normal-metal-insulator-superconducting junction and can remove photons incoherently from the system mediated by electron tunneling at specific bias-voltage pulses~\cite{Vadimov2022,Moerstedt2022}.}
    \label{fig:model}
\end{figure}

Our general model shown in Fig.~\subref*{fig:modela} consists of a system of $N$ harmonic modes and of an environment such that each system mode is interacting with their local Markovian bath. The $j$th mode is described by annihilation and creation operators $\ha_j$ and $\hdgg{a}_j$, respectively, with the canonical commutation relations $[\ha_j,\hdgg{a}_k]=\delta_{jk}$. We assume that the modes are linearly coupled to one another in any desired topology yielding up to quadratic terms in their coupling Hamiltonian. As an example realization of such a general model, we explore in Secs.~\ref{sec:modelb}--\ref{sec:resultsb} a linear chain of three lossy superconducting resonators as shown in Fig.~\subref*{fig:modelb}. Quadratic Hamiltonians can also be employed to accurately describe specific nonlinear systems, such as an optomechanical system subjected to a strong optical pump~\cite{Barzanjeh2021}. 

By defining the quadrature operators of the $j$:th mode as $\hq_j=(\ha_{j}+\hdgg{a}_j)/\sqrt{2}$ and $\hp_j=-i(\ha_{j}-\hdgg{a}_j)/\sqrt{2}$ and their $2N$-dimensional vector as $\mathbf{\hx}=(\hq_1,\hp_1,...,\hq_N,\hp_N)^{\top}$, the total Hermitian Hamiltonian describing the system classically driven by amplitudes $\mathbf{c}=(c_1,...,c_{2N})^{\top}$ can be cast into the compact quadratic form~\cite{Nicacio2017} 
\begin{align}
    \hH=\frac{1}{2}\mathbf{\hx}^{\top}\mathbf{H}\mathbf{\hx}+\mathbf{c}^{\top}\mathbf{\Omega}\mathbf{\hx},\label{eq:quadH}
\end{align}
where we dropped possible constant energy offsets, introduced the $2N\times2N$ symmetric matrix $\mathbf{H}$ carrying the internal and mode--mode coupling energies, and utilized the symplectic matrix,
\begin{align}
    \mathbf{\Omega}=\bigoplus_{j=1}^{N}\left(\begin{array}{cc}
0 & 1\\
-1 & 0
\end{array}\right).\label{eq:Jmat}
\end{align}
The commutation relations between the elements of $\mathbf{\hx}$ read $[\mathbf{\hx}_j,\mathbf{\hx}_k]=i\mathbf{\Omega}_{jk}$. Note that $\{\hq_j\}$ and $\{\hp_j\}$ play the role of generalized dimensionless position and momentum operators such that for superconducting $LC$ circuits they are related to flux and charge operators, respectively~\cite{Blais2021}.

For the sake of simplicity, we assume throughout this work that the system is only locally coupled to independent low-temperature environments. Consequently, after tracing out the environmental degrees of freedom, the temporal evolution of the reduced density operator of the system, $\hrh$, is given by the Lindblad master equation $\textrm{d}\hrh/\textrm{d}t=-i[\hH,\hrh]/\hbar+\mathcal{L}_{\downarrow}(\hrh)$, where
\begin{align}
    \mathcal{L}_\downarrow(\hrh)=\frac{1}{2\hbar}\sum_{j=1}^{N}\left[2\hL_{j}^{\downarrow}\hrh(\hL_{j}^{\downarrow})^{\dagger}-\left\{(\hL_{j}^{\downarrow})^{\dagger}\hL_{j}^{\downarrow},\hrh\right\} \right],\label{eq:LME}
\end{align}
describes the incoherent dynamics of the system associated to the jump operators $\{\hL_{j}^{\downarrow}\}$, each of which removes a photon from the corresponding mode. Namely, $\hL_{j}^{\downarrow}=\sqrt{\hbar\Gamma_{j}^{\downarrow}}\ha_j$, where $\{\Gamma_{j}^{\downarrow}\}$ are energy decay rates. In the above equations, there are no terms which induce thermal excitations since we assumed the baths to be at low temperatures.  
Note that $\hL_{j}^{\downarrow}$ can be written as a linear combination of the elements of $\mathbf{\hx}$, i.e., $\hL_{j}^{\downarrow}=(\mathbf{u}_{j}^{\downarrow})^{\top}\mathbf{\Omega}\mathbf{\hx}$, with coefficients given by a $2N$-dimensional vector $\mathbf{u}_{j}^{\downarrow}$.

Under the above conditions and for an initial Gaussian state of the $N$ oscillators, the dynamics of the system can be fully characterized by the so-called mean vector and covariance matrix (CM), the components of which are $\ev{\mathbf{\hx}_j}=\Tr(\mathbf{\hx}_j\hrh)$ and $\mathbf{V}_{jk}=\frac{1}{2}\left(\ev{\mathbf{\hx}_{j}\mathbf{\hx}_{k}}+\ev{\mathbf{\hx}_{k}\mathbf{\hx}_{j}}\right)-\ev{\mathbf{\hx}_{j}}\ev{\mathbf{\hx}_{k}}$, respectively. Here, we aim to solve the dynamics of the CM since it captures all squeezing and nonlocal properties of the system. By differentiating $\mathbf{V}$ with respect to time and using Eq.~\eqref{eq:LME}, we verify that the CM evolves according to the differential Lyapunov equation~\cite{Nicacio2015},
\begin{align}
    \frac{\textrm{d}\mathbf{V}}{\textrm{d}t}=\mathbf{\Gamma}\mathbf{V}+\mathbf{V}\mathbf{\Gamma}^{\top}+\mathbf{D},\label{eq:dyneq}
\end{align}
where we defined the $2N\times2N$ matrices $\mathbf{\Gamma}=\mathbf{\Omega}(\mathbf{H}-\textrm{Im}\mathbf{\Upsilon})/\hbar$, $\mathbf{D}=\textrm{Re}\mathbf{\Upsilon}/\hbar$, and $\mathbf{\Upsilon}=\sum_{j}[\mathbf{u}_{j}^{\downarrow}(\mathbf{u}_{j}^{\downarrow})^{\dagger}]$. The CM is a real, symmetric, and positive-definite matrix. As a compact statement of the uncertainty principle, the CM must also fulfill the condition $\mathbf{V}+i \mathbf{\Omega}/2\geq0$~\cite{Simon1994}. 

Below, we focus on the scenario where $\mathbf{\Gamma}$ and $\mathbf{D}$ are independent of time. Given an initial CM $\mathbf{V}(0)\equiv\mathbf{V}_0$, the solution of Eq.~\eqref{eq:dyneq} in this case is given by~\cite{Gajic1995}
\begin{align}
    \mathbf{V}(t)=\textrm{e}^{\mathbf{\Gamma}t}\left(\mathbf{V}_0 - \mathbf{V}_{\textrm{ss}}\right)\textrm{e}^{\mathbf{\Gamma}^{\top}t}+\mathbf{V}_{\textrm{ss}},\label{eq:Vsol}
\end{align}
where $\mathbf{V}_{\textrm{ss}}$ is the steady-state CM obtained as the solution of the algebraic Lyapunov equation $\mathbf{\Gamma}\mathbf{V}_{\textrm{ss}}+\mathbf{V}_{\textrm{ss}}\mathbf{\Gamma}^{\top}+\mathbf{D}=0$. 
We observe from Eqs.~\eqref{eq:dyneq} and~\eqref{eq:Vsol} that $\mathbf{\Gamma}$ has the role of a dynamical matrix so that all possible EPs are determined by its structure. 
Since the entries of $\mathbf{\Gamma}$ are real numbers with units of angular frequency, its eigenvalues are the complex-conjugate pairs $\lambda_{\mathbf{s}_m}^{\pm}$. Here, we define the index $\mathbf{s}_m=(m,\mu_m)$ to refer to the $m$th pair of the eigenvalues of $\mathbf{\Gamma}$, each eigenvalue having a multiplicity $\mu_m$. Observe that the maximum allowed multiplicity is, thus, $\max(\mu_m)=N$. 

The matrix $\mathbf{\Gamma}$ admits a Jordan normal form $\mathbf{\Gamma}=\mathbf{P}\mathbf{J}\mathbf{P}^{-1}$, where $\mathbf{P}$ is a nonsingular matrix and $\mathbf{J} = \text{diag}[\mathbf{J}_{\mathbf{s}_1}^{-}(\lambda_{\mathbf{s}_1}^{-}),...,\mathbf{J}_{\mathbf{s}_k}^{+}(\lambda_{\mathbf{s}_k}^{+})]$. The Jordan blocks $\mathbf{J}_{\mathbf{s}_m}^{\pm}(\lambda_{\mathbf{s}_m})$ can be decomposed as $\mu_m\times \mu_m$ matrices $\mathbf{J}_{\mathbf{s}_m}^{\pm}(\lambda_{\mathbf{s}_m}^{\pm})=\lambda_{\mathbf{s}_m}^{\pm} \mathbf{I}_{\mu_m}+\mathbf{N}_{\mu_m}$ with $\mathbf{I}_{\mu_m}$ being the identity matrix and $\mathbf{N}_{\mu_m}$ having the elements above the diagonal filled with ones. Naturally, the Jordan blocks for $\mu_m=1$ are just the scalars $\lambda_{\mathbf{s}_m}^{\pm}$. With these definitions, Eq.~\eqref{eq:Vsol} can be rewritten as
\begin{align}
    \mathbf{V}(t)=\mathbf{P}\textrm{e}^{\mathbf{J}t}\mathbf{P}^{-1}\left(\mathbf{V}_0 - \mathbf{V}_{\textrm{ss}}\right)\left(\mathbf{P}^{-1}\right)^{\top}\textrm{e}^{\mathbf{J}^{\top}t}\mathbf{P}^{\top}+\mathbf{V}_{\textrm{ss}},\label{eq:VsolJ}
\end{align}
where $\textrm{e}^{\mathbf{J}t}=\text{diag}(\textrm{e}^{\lambda_{\mathbf{s}_1}^{-} t}\textrm{e}^{\mathbf{N}_{\mu_1}t},...,\textrm{e}^{\lambda_{\mathbf{s}_k}^{+} t}\textrm{e}^{\mathbf{N}_{\mu_k}t})$.

The emergence of EPs and the associated critical dynamics of the CM correspond to the cases where the dynamical matrix $\mathbf{\Gamma}$ becomes nondiagonalizable, i.e., for any $\mu_m>1$. In other words, degeneracies in the spectrum of $\mathbf{\Gamma}$ produce nilpotent matrices $\mathbf{N}_{\mu_m}t$, the exponentials of which yield polynomials in time. Hereafter, these non-Hermitian degeneracies will be referred to as EP-$\mu_m$. Considering the definition of $\mathbf{\Gamma}$, we remark that the term $\mathbf{\Omega}\mathbf{H}$ itself does not promote critical dynamics as it gives rise to unitary evolution of the CM. The production of EPs must be accompanied with the incoherent processes caused by the local environments and attributed to the term $\mathbf{\Omega}\textrm{Im}\mathbf{\Upsilon}$.

To summarize, Eq.~\eqref{eq:VsolJ} is valid for any time-independent matrices $\mathbf{\Gamma}$ and $\mathbf{D}$ describing the evolution of a system of coupled quantum harmonic oscillators in noisy Gaussian channels yielding the steady-state CM $\mathbf{V}_{\textrm{ss}}$. At an EP, Eq.~\eqref{eq:VsolJ} reveals that the solution linked to the critical dynamics is an exponential function multiplied by a polynomial, which will be explored below in specific cases. Alternatively, the description of EPs for quadratic Liouvillians, such as the one related to Eq.~\eqref{eq:LME}, may be given in terms of annihilation and creation operators as recently developed in Ref.~\cite{Arkhipov2021}.

\section{Three coupled resonators under individual losses}\label{sec:modelb}

The system and its environment considered in this paper is depicted in Fig.~\subref*{fig:modelb}. Three superconducting resonators, $R1$, $R2$, and $R3$ are capacitively coupled in a linear-chain configuration through a fixed coupling constant $g>0$. We focus on a single electromagnetic mode for each resonator, which, including the coherent couplings, defines our system. Each mode may dissipate its energy into its independent linear bath. Nevertheless, quantum effects may emerge at low temperatures and for sufficiently high quality factors and for nonclassical initial states~\cite{Blais2021}, and, consequently, we need to employ a quantum-mechanical model. 

In the single-mode and rotating-wave approximations, the Hamiltonian of the system reads
\begin{equation}
\hH=\hbar\sum_{j=1}^{3}\omega_j\left(\hdgg a_{j}\ha_{j}+\frac{1}{2}\right)+\hbar g(\ha_{1}\hdgg a_{2}+\ha_{2}\hdgg a_{3}+\text{H.c.}),\label{eq:H3}
\end{equation}
where $\omega_j$ is the fundamental angular frequency of the $j$th resonator, $\{\ha_j\}$ are the corresponding ladder operators defined as in Sec.~\ref{sec:modela}, and H.c. refers to the Hermitian conjugate. The losses of the system are modeled here as in Eq.~\eqref{eq:LME} with jump operators $\hL_j^{\downarrow}=\sqrt{\hbar \kappa_j}\ha_j$ and decay rates $\kappa_j>0$ for $j=1$--$3$. Some of the decay rates can be adjusted experimentally through the QCRs shown in Fig.~\subref*{fig:modelb}. As we show below, to produce EP-$3$ with degenerate resonators, we need asymmetric decay rates, a scenario, which can be realized by the two independent QCRs shown in Fig.~\subref*{fig:modelb}. 

By writing the ladder operators in terms of the quadrature operators as $\ha_j=(\hq_j+i\hp_j)/\sqrt{2}$ and using the notation of Sec.~\ref{sec:modela}, the $6\times6$ dynamical matrix $\mathbf{\Gamma}$ becomes 

\begin{align}
\mathbf{\Gamma}=\left(\begin{array}{ccc}
\mathbf{K}_{1} & \mathbf{G} & \mathbf{0}_2\\
\mathbf{G} & \mathbf{K}_{2} & \mathbf{G}\\
\mathbf{0}_2 & \mathbf{G} & \mathbf{K}_{3}
\end{array}\right),\label{eq:Gamma}
\end{align}
where $\mathbf{0}_2$ is the $2\times 2$ null matrix and
\begin{align}
\mathbf{K}_{j}=\left(\begin{array}{cc}
-\frac{\kappa_{j}}{2} & \omega_{j}\\
-\omega_{j} & -\frac{\kappa_{j}}{2}
\end{array}\right),\ \ \ 
\mathbf{G}=\left(\begin{array}{cc}
0 & g\\
-g & 0
\end{array}\right).\label{eq:Gammakg}
\end{align}
By denoting the single-mode CM of the vacuum state as $\mathbf{V}^{(j)}_{\text{vac}}=\text{diag}\left(1,1\right)/2$, one readily obtains
\begin{align}
\mathbf{D}&=\bigoplus_{j=1}^{3}\kappa_{j}\mathbf{V}^{(j)}_{\text{vac}},\  \mathbf{V}_{\text{ss}}=\bigoplus_{j=1}^{3}\mathbf{V}^{(j)}_{\text{vac}},\label{eq:Dmat}
\end{align}
the latter corresponding to the CM of any product of three coherent states.
Since the jump operators here do not promote incoherent displacements, the steady state is actually the three-mode vacuum state $\ket{0}_1\ket{0}_2\ket{0}_3$ as long as all $\kappa_j>0$. 

\subsection{Characterization of exceptional points}\label{sec:EPs}
Finding the EPs directly from the spectrum of $\mathbf{\Gamma}$ may be challenging as one needs to solve a $2N$th degree polynomial equation, or in the studied case, a sextic equation. However, owing to the absence of counter-rotating terms in the form of $\hH$, here, the characterization of EPs can be simplified to the study of the dynamical equation for the $3\times3$ vector $\mathbf{a}_{\text{rf}}=(\ev{\ha_1}_{\text{rf}},\ev{\ha_2}_{\text{rf}},\ev{\ha_3}_{\text{rf}})^{\top}$, where $\ev{\ha_j}_\text{rf}$ are expectation values calculated at a frame rotating at angular frequency $\omega_1$ about $\sum_{j=1}^3\hat{a}_j^\dagger \hat{a}_j$, see Appendix~\ref{app:EPs}. In such a frame, one can obtain $\dot{\mathbf{a}}_{\text{rf}}=-i\mathcal{H}\mathbf{a}_{\text{rf}}$, with $\mathcal{H}$ having the role of an effective non-Hermitian Hamiltonian. Explicitly, we have
\begin{align}
    \mathcal{H}=\left(\begin{array}{ccc}
-i\frac{\kappa_{1}}{2} & g & 0\\
g & \delta_2-i\frac{\kappa_{2}}{2} & g\\
0 & g & \delta_3-i\frac{\kappa_{3}}{2}
\end{array}\right),\label{eq:nH2}
\end{align}
where $\delta_2=\omega_2-\omega_1$ and $\delta_3=\omega_3-\omega_1$ are frequency detunings. 

\begin{figure}
    \subfloat{\label{fig:epsa}} 
	\subfloat{\label{fig:epsb}}
	\subfloat{\label{fig:epsc}}
	\subfloat{\label{fig:epsd}}
    \centering
    \includegraphics[width=\linewidth]{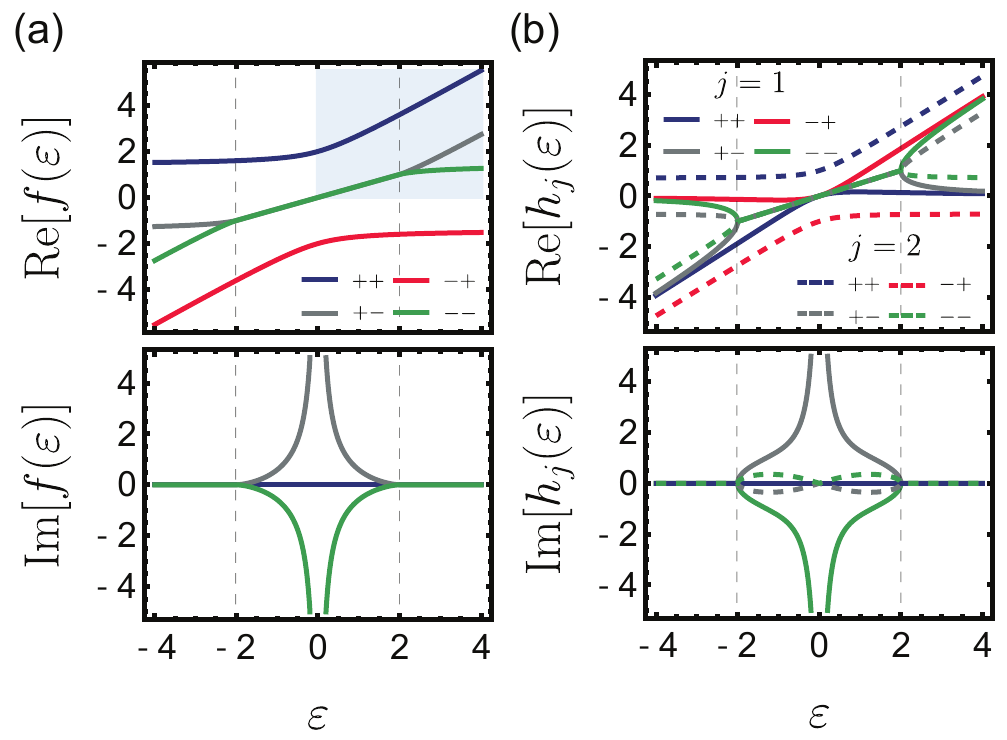}
    \caption{Exceptional-point engineering for a linear chain of three lossy resonators with degenerate angular frequencies $\omega_1=\omega_3$, expressed by a purely real parameter $\varepsilon=\varepsilon_3$, see Eqs.~\eqref{eq:deltakappa}--\eqref{eq:hrealeps}. (a) Decay rate (top panel) and frequency (bottom panel) offsets of resonator $R2$ as functions of the decay rate offset of resonator $R3$, expressed by the complex-valued function $f(\varepsilon)=\varepsilon_2$ defined in Eq.~\eqref{eq:frealeps}. (b) Effective decay rate (top) and effective frequency (bottom) offsets of the eigenmodes of the system as functions of the decay rate offset of resonator $R3$, expressed by the complex-valued functions $h_j(\varepsilon)$ defined in Eq.~\eqref{eq:hrealeps}. All offsets are given with respect to the parameters of resonator $R1$. In (b), solid (dashed) curves represent the single (double) roots of the characteristic polynomial of $\mathcal{H}$. In all cases, the labels $++$, $-+$, $+-$, and $--$ indicate the four branches of $f(\varepsilon)$ obtained from the corresponding selection of signs in Eq.~\eqref{eq:frealeps}. The vertical dashed lines in all panels highlight the values of $\varepsilon$ producing EP-$3$. The shaded area in (a) indicates the relevant region of the $\text{Re}(\varepsilon_3)$-$\text{Re}(\varepsilon_2)$ parameter space for this paper.}
    \label{fig:eps}
\end{figure}

Without loss of generality, we assume that the parameters $g$, $\omega_1$, and $\kappa_1$ are fixed. Thus, it is convenient to express the parameters of $R2$ and $R3$ with respect to those of $R1$. We proceed with this parametrization using complex-valued parameters \{$\varepsilon_k$\} such that for $k=2,3$, we have
\begin{align}
   \delta_{k}(\varepsilon_k)=\sqrt{2}g\text{Im}(\varepsilon_k), \ \ \kappa_{k}(\varepsilon_k)=\kappa_{1}+2\sqrt{2}g\text{Re}(\varepsilon_k).\label{eq:deltakappa} 
\end{align}
As detailed in Appendix~\ref{app:EPs}, degeneracies in the spectrum of $\mathcal{H}$ appear provided that the relationship between $\varepsilon_2$ and $\varepsilon_3$ is expressed through the complex-valued function, 
\begin{align}
f(\varepsilon)=&\ \frac{1}{2}\left[\varepsilon\pm\sqrt{\frac{\varepsilon^{4}+10\varepsilon^{2}-2\pm2\left(1+2\varepsilon^{2}\right)^{\frac{3}{2}}}{\varepsilon^2}}\right],
\label{eq:frealeps}
\end{align}
where we defined $\varepsilon=\varepsilon_3$ and $f(\varepsilon)=\varepsilon_2$ to highlight the dependence of the parameters of $R2$ on those of $R3$ to produce EPs. Note that $f(\varepsilon)$ presents four branches indicated by the signs ``$\pm$'' as shown for a purely real $\varepsilon$ in Fig.~\subref*{fig:epsa}. 

At the degeneracies of $\mathcal{H}$, such a matrix has, at most, two distinct eigenvalues $\delta_{j}^{\text{eff}}(\varepsilon)-i\kappa_{j}^{\text{eff}}(\varepsilon)/2$ from which the effective detunings and decay rates of the normal modes are extracted as $\delta_{j}^{\text{eff}}(\varepsilon)=\sqrt{2}g\text{Im}[h_{j}(\varepsilon)]$ and $\kappa_{j}^{\text{eff}}(\varepsilon)=\kappa_{1}+2\sqrt{2}g\text{Re}[h_{j}(\varepsilon)]$ (Appendix~\ref{app:EPs}), where 
\begin{align}
h_{1}(\varepsilon)&=\frac{f^{3}-\varepsilon f^{2}-(\varepsilon^{2}+4)f+\varepsilon^{3}+\varepsilon/2}{f^{2}-\varepsilon f+\varepsilon^{2}-3},\nonumber\\
h_{2}(\varepsilon)&=h_{3}(\varepsilon)=\frac{1}{4}\left[\frac{2\varepsilon f^{2}+2(\varepsilon^{2}+1)f-7\varepsilon}{f^{2}-\varepsilon f+\varepsilon^{2}-3}\right],
\label{eq:hrealeps}
\end{align}
and we write $f=f(\varepsilon)$ for brevity. Consequently, the degenerate eigenvalues of $\mathbf{\Gamma}$ are given by the pairs (Appendix~\ref{app:EPs}), 
\begin{equation}
    \lambda_{\mathbf{s}_{j}}^{\pm}(\varepsilon)=-\frac{\kappa_{j}^{\text{eff}}(\varepsilon)}{2}\pm i\left[\omega_{1}+\delta_{j}^{\text{eff}}(\varepsilon)\right],\label{eq:lambdaEP}
\end{equation}
which coincide at an EP-$3$. The rich structure of the decay rates and frequencies of the normal modes is shown in Fig.~\subref*{fig:epsb} for a purely real $\varepsilon$.

Without imposing further restrictions, the considered open system presents six EP-$3$'s, two of which are obtained for $\varepsilon=2f(\varepsilon)=\pm2$ so that all modes are degenerate, $\kappa_2=\kappa_1\pm2\sqrt{2}g$ and $\kappa_3=\kappa_1\pm4\sqrt{2}g$. These cases correspond to the square-root singularity of $f(\varepsilon)$ and are highlighted in Fig.~\ref{fig:eps}. The remaining four EP--$3$ are obtained with $f(\varepsilon)=(\pm3\sqrt{3}\pm i)/(2\sqrt{2})$, and $\varepsilon=2i\Im[f(\varepsilon)]=\pm i/\sqrt{2}$, thus, requiring equal decay rates for $R1$ and $R3$, $\kappa_2=\kappa_1\pm3\sqrt{3}g$, in addition to the detunings $\delta_2=\pm g/2$ and $\delta_3=\pm g$. The degeneracy map for such cases is shown in Fig.~\ref{fig:ndeps} of Appendix~\ref{app:EPs} for completeness. 

All other cases expressed through Eqs.~\eqref{eq:frealeps} and~\eqref{eq:hrealeps} are associated with EP-$2$. Our numerical tests show the coalescence of eigenvectors of $\mathcal{H}$ following the branches $f(\varepsilon)$, indeed indicating open-system degeneracies. The Jordan decompositions of $\mathbf{\Gamma}$ yielding polynomial-in-time features of the dynamics are shown in Appendix~\ref{app:Jordan} for relevant EPs in this paper.

We emphasize that the experimental feasibility of EP engineering in the present model is strongly dependent on the physical limitations of the setup. For instance, to obtain the four instances of EP-$3$ with nondegenerate frequencies, one needs frequency detunings on the order of $g/(2\pi)$, which are typically much smaller than the frequency of superconducting resonators themselves~\cite{Blais2021}. Hereafter, we restrict our discussion to degenerate resonators, i.e., $\Im(\varepsilon)=\Im[f(\varepsilon)]=0$. By also considering $\kappa_1$ as the smallest decay rate, another restriction for obtaining EPs is imposed such that both $\text{Re}(\varepsilon)\geq 0$ and $\text{Re}[f(\varepsilon)]\geq 0$. In this case, the only allowed branches of $f(\varepsilon)$ are ``$+-$'' and ``$--$'' for $\varepsilon\geq2$, and ``$++$'' for $\varepsilon\geq0$, see the shaded region in Fig.~\subref*{fig:epsa}. In particular, the branch $++$ at $\varepsilon=0$ yields weakly dissipative normal modes, with one of them decaying according to $\kappa_1$, see Fig.~\subref*{fig:epsb} and Eq.~\eqref{eq:hrealeps}. This behavior suggests that a quasistabilization of some properties of the system can be obtained with the combination of a small $\kappa_1$ and a proper choice of the EP as explored in detail in Sec.~\ref{sec:resultsa}.

\subsection{Single-mode squeezing and bipartite entanglement}

Below, we specifically investigate single-mode squeezing and bipartite entanglement for the three-resonator system. For Gaussian evolution, these quantities can be addressed directly from the specific partitions of the total CM,
\begin{align}
\mathbf{V}=\left(\begin{array}{ccc}
\mathbf{V}^{(1)} & \mathbf{C}^{(12)} & \mathbf{C}^{(13)}\\
\mathbf{C}^{(12)\top} & \mathbf{V}^{(2)} & \mathbf{C}^{(23)}\\
\mathbf{C}^{(13)\top} & \mathbf{C}^{(23)\top} & \mathbf{V}^{(3)}
\end{array}\right),\label{eq:V6x6}
\end{align}
where $\mathbf{V}^{(j)}$ is the reduced CM of resonator $Rj$ and $\mathbf{C}^{(jk)}$ is the intermodal correlation matrix between resonators $Rj$ and $Rk$~\cite{Nicacio2017}.

Since all single-mode Gaussian states can be written as squeezed thermal states apart from local displacements, the components of the reduced CM of resonator $Rj$ can be cast into the form~\cite{Ferraro2005}
\begin{align}
    \mathbf{V}_{11}^{(j)}&=(\bar{N}_j+1/2)[\cosh(2r_j)+\sinh(2r_j)\cos\phi_j],\nonumber\\
    \mathbf{V}_{22}^{(j)}&=(\bar{N}_j+1/2)[\cosh(2r_j)-\sinh(2r_j)\cos\phi_j],\nonumber\\
    \mathbf{V}_{12}^{(j)}&=(\bar{N}_j+1/2)\sinh(2r_j)\sin\phi_j,
\end{align}
where $r_j$ and $\phi_j$ are real-valued quantities defining the squeezing parameter $\xi_j=r_je^{i\phi_j}$ and $\bar{N}_j$ is the effective thermal occupation number of resonator $Rj$. As a consequence, one can extract $r_j$ and $\bar{N}_j$ as
\begin{align}
    r_j&=\frac{1}{2}\sinh^{-1}\left[\frac{\sqrt{(\mathbf{V}_{11}^{(j)}-\mathbf{V}_{22}^{(j)})^2+4\mathbf{V}_{12}^{(j)2}}}{2(\bar{N}_j+1/2)}\right],\nonumber\\
    \bar{N}_j&=\sqrt{\det\mathbf{V}^{(j)}}-\frac{1}{2},\label{eq:param}
\end{align}
and the single-mode purity is readily given by $\mathcal{P}_j=(2\bar{N}_j+1)^{-1}$.

Although bipartite entanglement can be quantified by the reduced von Neuman entropy given a pure state of the complete system~\cite{Nielsen2000}, an entanglement measure for the mixed states is not uniquely defined~\cite{Horodecki2009}. Here, we focus on the concept of logarithmic negativity~\cite{Vidal2002}, which is based on the Peres-Horodecki separability criterion~\cite{Peres1996,Simon2000} and fulfills the conditions for an entanglement monotone~\cite{Plenio2005a}. 

Given Eq.~\eqref{eq:V6x6} and considering the subsystems $Rj$ and $Rk$ ($j<k$), one can write their joint CM as
\begin{align}
\mathbf{V}^{(jk)}=\left(\begin{array}{cc}
\mathbf{V}^{(j)} & \mathbf{C}^{(jk)}\\
\mathbf{C}^{(jk)\top} & \mathbf{V}^{(k)}
\end{array}\right).\label{eq:CM2}
\end{align}
For Gaussian states, the logarithmic negativity $\mathcal{E}_{jk}$ can then be computed as~\cite{Vidal2002} 
\begin{align}
    \mathcal{E}_{jk}=\max[0,-\log_2(2\tilde{\nu}_{jk}^{-})],\label{eq:logneg}
\end{align}
where $\tilde{\nu}_{jk}^{-}=\{\tilde{\Delta}_{jk}-[\tilde{\Delta}_{jk}^2-4\det\mathbf{V}^{(jk)}]^{\frac{1}{2}}\}^{\frac{1}{2}}/\sqrt{2}$ being the smallest symplectic eigenvalue of $\mathbf{\tilde{V}}^{(jk)}$, which corresponds to the two-mode CM obtained after the Peres-Horodecki partial transposition of the associated bipartite density matrix, and $\tilde{\Delta}_{jk}=\det\mathbf{V}^{(j)}+\det\mathbf{V}^{(k)}-2\det\mathbf{C}^{(jk)}$~\cite{Simon2000,Adesso2004}. The inequality $\tilde{\nu}_{jk}^{-}\geq1/2$ is a necessary and sufficient condition for separability of 
bipartite Gaussian systems of two modes~\cite{Simon2000,Adesso2004}.

\section{Quasistabilization of squeezing and entanglement}\label{sec:resultsa}

In this section, we study the propagation of single-mode squeezing and bipartite entanglement in the open quantum system of Fig.~\subref*{fig:modelb}. The initial state is chosen as $\ket{0}_1\ket{0}_2\hS_3(r)\ket{0}_3$, where $\hS_3(r)=\exp[r(\hat{a}^{\dagger 2}_3-\hat{a}^{2}_3)/2]$ is the single-mode squeezing operator of $R3$ and $r\geq0$. Such a state has the CM,
\begin{align}
    \mathbf{V}_{0}&=\frac{1}{2}\text{diag}\left(1,1,1,1,\text{e}^{2r},\text{e}^{-2r}\right),\label{eq:V0}
\end{align}
which indicates that the variances of $R3$ are initially modified by the factors $\text{e}^{\pm2r}$.
We employ Eq.~\eqref{eq:Vsol} to numerically obtain the $6\times6$ time-evolved CM $\mathbf{V}(t)$ at different points of the parameter space. Here, we set $\kappa_1=\kappa_3$ as the smallest decay rates of the system and test different $\kappa_2=\kappa_1+2\sqrt{2}g\Re(\varepsilon_2)$ with $\Re(\varepsilon_2)\geq0$ and $\Im(\varepsilon_2)=0$. Within these conditions, the only allowed EP-branch is $++$ so that an EP-$2$ is produced at $f(\varepsilon=0)=2$, see Eq.~\eqref{eq:frealeps} and Fig.~\subref*{fig:epsa}. 

\begin{figure}
    \centering
    \subfloat{\label{fig:dyna}} 
	\subfloat{\label{fig:dynb}}
    \includegraphics[width=\linewidth]{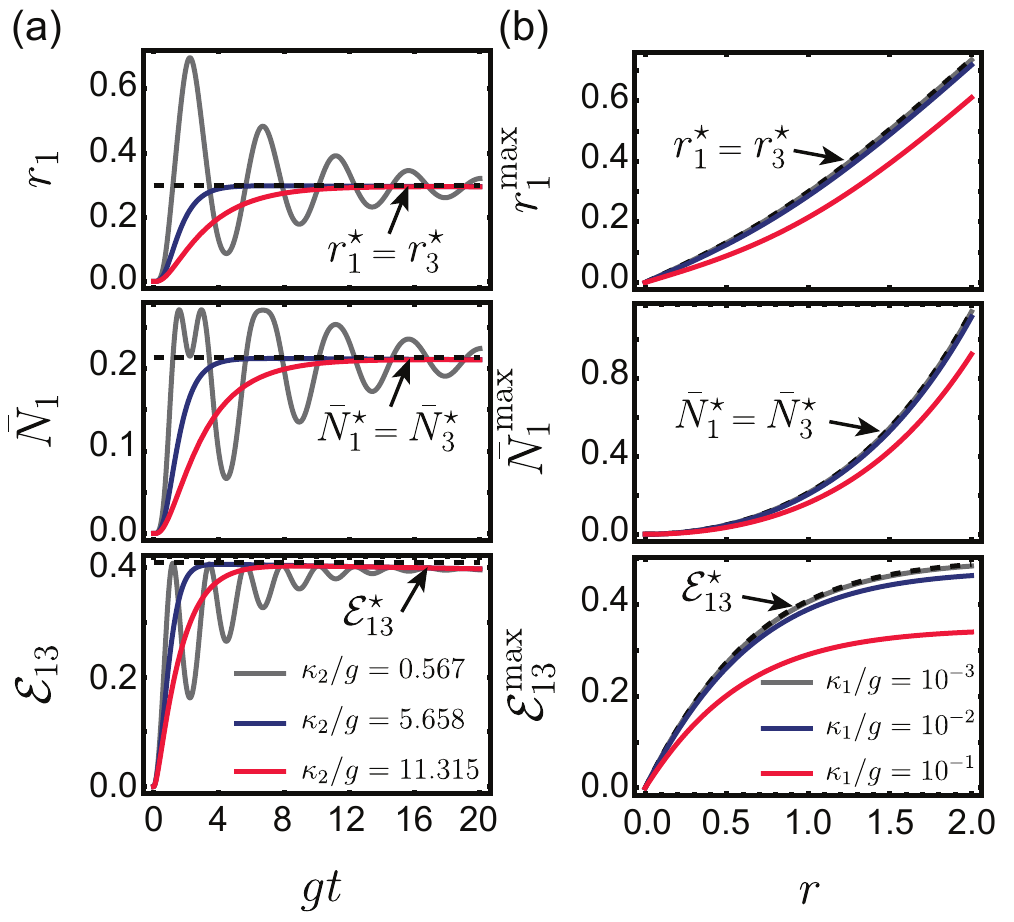}
    \caption{(a) Dynamics of the squeezing parameter $r_1$ and effective thermal occupation $\bar{N}_1$ of resonator $R1$ and the logarithmic negativity between $R1$ and $R3$, $\mathcal{E}_{13}$ for the indicated values of the damping rate of $R2$, $\kappa_2/g$. The shown data correspond to a crossover from underdamped to overdamped dynamics with critical damping at $\kappa_2/g=5.658$. The frequencies of the resonator modes are chosen as $\omega_1/g=\omega_2/g=\omega_3/g=5000$, the other damping rates as $\kappa_1/g=\kappa_3/g=10^{-3}$, and the initial squeezing parameter of $R3$ as $r=1$. The corresponding values of $\varepsilon_2$ as defined in Eq.~\eqref{eq:deltakappa} are $\varepsilon_2=0.2$ (gray curves), $\varepsilon_2=2.0$ (blue curves), and $\varepsilon_2=4.0$ (red curves).
    In the chosen parameter regime, the results are essentially independent of the resonator-resonator coupling strength $g$. (b) Maximum achieved quantities in temporal evolutions corresponding to (a) at the critical damping as functions of the initial squeezing parameter $r$ for selected values of $\kappa_1/g$. In all panels, dashed lines correspond to long-time values in the limit $\kappa_1/g\rightarrow0$, see Eq.~\eqref{eq:paramax}.}
    \label{fig:dyn}
\end{figure}

In Figure~\subref*{fig:dyna}, we observe the emergence of squeezed thermal states for resonator $R1$ and bipartite quantum correlations expressed through the logarithmic negativity $\mathcal{E}_{13}$ with a clear passage from underdamped to overdamped dynamics with increasing $\kappa_2$. The squeezing degree of $R2$ along with the logarithmic negativities $\mathcal{E}_{12}$ and $\mathcal{E}_{23}$ (data not shown) is rapidly suppressed for large ratios $\kappa_2/g$. On the other hand, the small values of $\kappa_j/g$, $j=1,3$, help to delay the decay of the system towards the three-mode vacuum state, and this quasistability tends to be achieved faster near the critical-damping regime produced by the EP-$2$. Such a behavior is not present at the EP-$2$ if $R1$ is directly connected to $R3$, which reduces the dimension of the system to $N=2$. In such a case, the only two normal modes of the system decay at equal rates~\cite{Partanen2019}.  

The maximum achieved values of $r_1$, $\bar{N}_1$, and $\mathcal{E}_{13}$ as functions of the initial squeezing parameter $r$ for the system dynamics at the EP-$2$ are shown in Fig.~\subref*{fig:dynb}. Their values in the limit $\kappa_j\to 0$, $j=1,3$, and $t\to\infty$, can be estimated directly from Eqs.~\eqref{eq:param} and~\eqref{eq:logneg} with the help of the Jordan decomposition of $\mathbf{\Gamma}$ shown in Appendix~\ref{app:Jordan}. One readily obtains $r_2^{\star}=\bar{N}_2^{\star}=\mathcal{E}_{12}^{\star}=\mathcal{E}_{23}^{\star}=0$, whereas,
\begin{align}
    r_1^{\star}=r_3^{\star}&=\frac{1}{2}\log\left[\frac{3+\text{e}^{2r}}{\sqrt{10+6\cosh(2r)}}\right],\nonumber\\
    \bar{N}_1^{\star}=\bar{N}_3^{\star}&=\frac{1}{8}\left[\sqrt{10+6\cosh(2r)}-4\right],\nonumber\\
    \mathcal{E}_{13}^{\star}&=\frac{1}{2}\left[1-\log_2(1+\text{e}^{-2r})\right].\label{eq:paramax}
\end{align}
The superscripts ``$\star$'' in Eqs.~\eqref{eq:paramax} indicate that such quantities are bounds for the quasistabilized states, shown as dashed lines in Fig.~\ref{fig:dyn}. Interestingly, we can generate entanglement between resonators $R1$ and $R3$ although the entanglement with resonator $R2$ is rapidly suppressed.

From Fig.~\subref*{fig:dynb} and Eqs.~\eqref{eq:paramax}, we observe that the squeezing splitting increases linearly with $r$ for $r\ll1$ where thermal occupancy is insignificant. The squeezing-splitting capacity $r_1^{\star}/r$ and the degree of entanglement between $R1$ and $R3$ tend to saturate to $1/2$ in the limit $r\rightarrow\infty$ with the expense of also thermally populating these resonators. Using the decibel scale defined by $r= 10\log_{10}(e^{2r})$~dB~\cite{Adesso2014}, an initial amount of squeezing $r\approx3$ dB is roughly converted into squeezed states with $r_1^{\star}=r_3^{\star}\approx0.772$ dB and purities $\mathcal{P}_1^{\star}=\mathcal{P}_3^{\star}\approx0.997$ with $\mathcal{E}_{13}^{\star}\approx0.207$. Despite producing a faster decay towards the actual steady state of the system, an increase in two orders of magnitude in $\kappa_1/g$ does not provide significant differences in the maximum quantities for small $r$.

\begin{figure}
\centering
\subfloat{\label{fig:staba}} 
\subfloat{\label{fig:stabb}}
\subfloat{\label{fig:stabc}}
\includegraphics[width=0.95\linewidth]{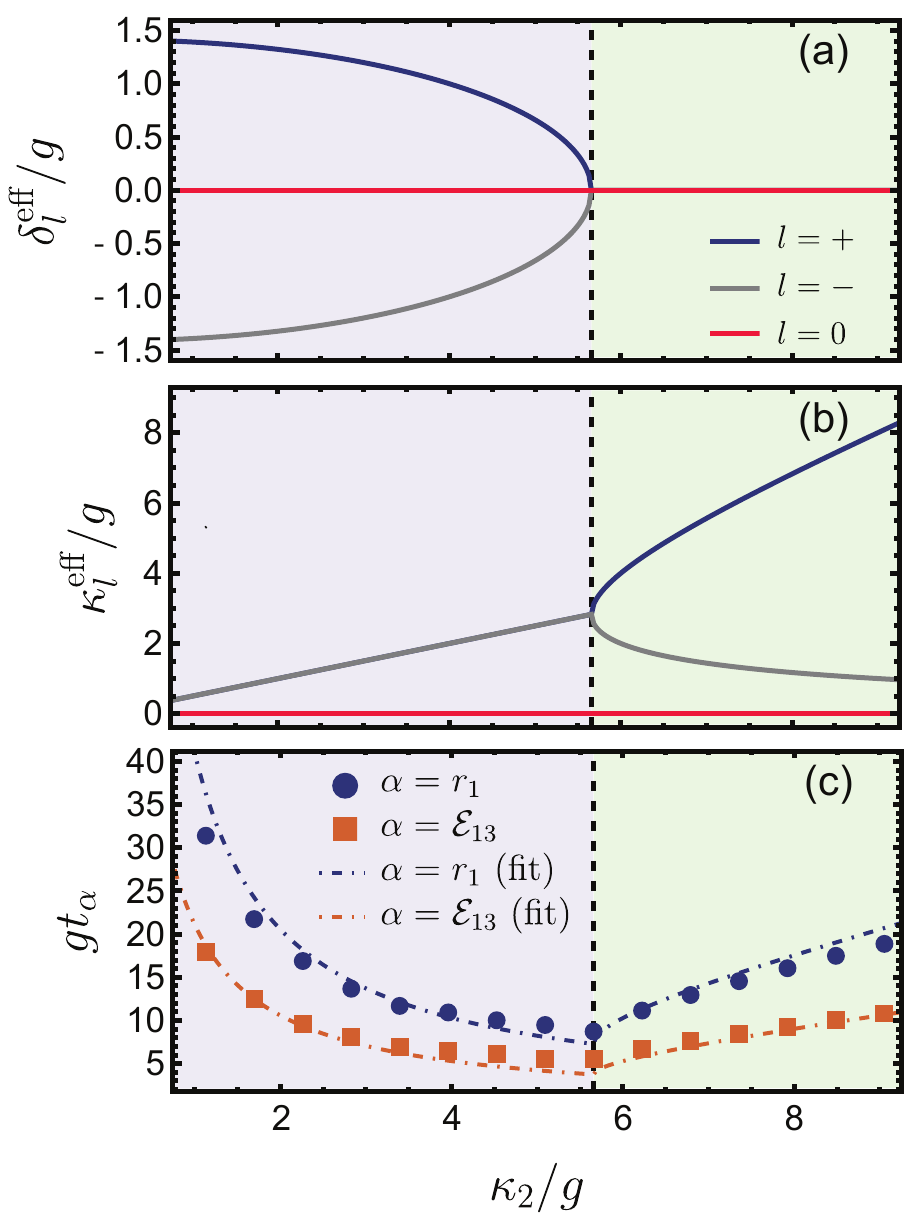}
\label{fig:stab}
\caption{(a) Effective frequency detunings and (b) effective decay rates of the eigenmodes of the coupled system as functions of the decay rate of resonator $R2$, $\kappa_2$, in units of the resonator-resonator coupling strength $g$. (c) Time $t_{\alpha}$ to yield quasistable squeezing (filled circles, $\alpha=r_1$) and entanglement (filled squares, $\alpha=\mathcal{E}_{13}$) within an uncertainty $\sigma_{\alpha}=10^{-5}$, see the main text. The dashed lines represent corresponding results from the fit functions of Eq.~\eqref{eq:fit}. In all panels, the parameters are chosen as in Fig.~\ref{fig:dyna}, and the colored regions separate the underdamped from the overdamped dynamics with critical damping at $\kappa_2/g=5.658$, corresponding to an EP-$2$.}
\end{figure}

To further address the quasistabilization of entanglement and squeezing transferred to $R1$ for different $\kappa_2$'s, we diagonalize Eq.~\eqref{eq:nH2} to obtain the effective frequency detunings and decay rates of the system as shown in Figs.~\subref*{fig:staba} and~\subref*{fig:stabb}, respectively. For $\kappa_1\ll\kappa_2$, we obtain two eigenmodes with frequency detunings $\delta_{\pm}^{\text{eff}}\approx\pm\Im(\sqrt{\kappa_2^2-32g^2})/4$ and dissipation rates $\kappa_{\pm}^{\text{eff}}\approx\kappa_2/2\pm\Re(\sqrt{\kappa_2^2-32g^2})/2$, which coalesce at $\kappa_2\approx4\sqrt{2}g$. The frequency detuning $\delta_0^{\text{eff}}=0$ and dissipation rate $\kappa_0^{\text{eff}}=\kappa_1$ are preserved, thus, indicating that one of the eigenmodes remains hidden from the dissipation of resonator $R2$. 

Since clearly the speed of quasistabilization for the squeezing and entanglement of resonator $R1$ depend on $\kappa_2$ [Fig.~\subref*{fig:dyna}] and since $\kappa_+^{\text{eff}}\geq \kappa_-^{\text{eff}}$, we conclude that the time scale for this quasi-stabilization is roughly given by $1/\kappa_-^{\text{eff}}\approx 2/[\kappa_2-\Re(\sqrt{\kappa_2^2-32g^2})]$. To arrive at a more accurate expression for the quasistabilization time, we first fit functions of the form 
\begin{align}
    r_1^{\text{fit}}(t)&=\frac{r_1^{\star}}{2}\text{e}^{-y_{r_1}\kappa_1 t}\left\{\text{e}^{- \kappa_{-}^{\text{eff}}t/2}\left[1-3\cos(\delta_{-}^{\text{eff}} t)\right]+2\right\},\nonumber\\
    \mathcal{E}_{13}^{\text{fit}}(t)&=\mathcal{E}_{13}^{\star}\text{e}^{-y_{\mathcal{E}_{13}}\kappa_1 t}\left[1-\text{e}^{- \kappa_{-}^{\text{eff}}t}\cos^2(\delta_{-}^{\text{eff}} t)\right],\label{eq:fit}    
\end{align}
to time traces similar to those in Fig.~\subref*{fig:dyna} and find $y_{r_1}\approx0.75$ and $y_{\mathcal{E}_{13}}\approx1.3$. Although these functions neglect the polynomial-in-time solution at the EP-$2$, they capture the main features of the over and underdamped dynamics, and, hence, are accurate enough from our following analysis.

Next, we define the quasistabilization time $t_\alpha$ as the earliest time instant after which the quantity $\alpha=r_1,\mathcal{E}_{13}$ stays within an uncertainty $\sigma_{\alpha}$ from the ideal value $\alpha^{\star}\text{e}^{-y_{\alpha}\kappa_1 t_{\alpha}}$ where we take into account also the slow decay of the maximum attainable value owing to finite $\kappa_1$. More precisely,  
\begin{align}
t_{\alpha} = \min\{t|\alpha^{\star}\text{e}^{-y_{\alpha}\kappa_1 t}-\tilde{\alpha}(t)\leq\sigma_{\alpha}\},\label{eq:error_new}
\end{align}
where $\tilde{\alpha}(t)$ is the lower envelope of the possibly oscillating $\alpha(t)$. Note that by this definition, $\tilde{\alpha}(t)=\alpha(t)$ in the critically and overdamped dynamics.

In Fig.~\subref*{fig:stabc}, we show the behavior of the quasistabilitation time $t_{\alpha}$ on the dissipation rates $\kappa_2$ for an error $\sigma_{\alpha}=10^{-5}$ as obtained from the solutions of the temporal evolution of the system similar to those in Fig.~\subref*{fig:dyna}. 
The shortest quasistabilization times are obtained in the vicinity of the EP-$2$ owing to the peak in $\kappa_{-}^{\text{eff}}$ illustrated in Fiq.~\subref*{fig:stabb}. Using the lower envelopes of the fitting functions~\eqref{eq:fit} in Eq.~\eqref{eq:error_new}, one can estimate the quasistabilization time as
\begin{align}
    t_{\alpha}\approx \frac{\log\left(\frac{\alpha^{\star}}{\sigma_{\alpha}}\right)}{y_{\alpha}\kappa_1+z_{\alpha}\kappa_{-}^{\text{eff}}},\label{eq:qstime}
\end{align}
with $z_{r_1}\approx0.5$ and $z_{\mathcal{E}_{13}}\approx1$. Therefore, $t_{\alpha}$ tends to scale logarithmically with the desired error.

\section{Fast reset near exceptional points}\label{sec:resultsb}

As the final application of EPs, we discuss the reset of the resonator chain to its ground-state $\ket{0}_1\ket{0}_2\ket{0}_3$. Typically, stronger dissipation leads to faster decay, but, of course, in our system where the coupling between the different resonators is weak compared with the excitation frequencies of the bare resonators, the critical dynamics plays an important role. Similar features are prone to arise in a quantum register of several coupled qubits.

To quantitatively study the accuracy of the reset, we define the infidelity, 
\begin{align}
    \mathcal{I}_{\text{ss}}(\hat\rho)=1-\mathcal{F}_{\text{ss}}(\hat\rho),\label{eq:Iss}
\end{align}
where $\mathcal{F}_{\text{ss}}(\hat\rho)=\bra{0}_1\bra{0}_2\bra{0}_3\hrh\ket{0}_1\ket{0}_2\ket{0}_3$ is the overlap probability between an arbitrary three-mode state $\hrh$ and the ground state. For multimode Gaussian states with the null mean vector $\ev{\mathbf{\hx}}$, $\mathcal{F}_{\text{ss}}$ can be directly computed from the covariance matrix $\mathbf{V}$, which for the present case becomes~\cite{Serafini2017}
\begin{align}
    \mathcal{F}_{\text{ss}}= \frac{1}{\sqrt{\det\left(\mathbf{V}+\mathbf{V}_{\text{ss}}\right)}},\label{eq:Fss}
\end{align}
where $\mathbf{V}_{\text{ss}}$ is given in Eq.~\eqref{eq:Dmat}. An optimized reset is achieved with the set of free parameters producing the fastest decay to the ground state, i.e., the minimal $\mathcal{I}_{\text{ss}}$ in a given time. 
\begin{figure}
    \subfloat{\label{fig:reseta}} 
	\subfloat{\label{fig:resetb}}
    \subfloat{\label{fig:resetc}}
    \centering
    \includegraphics[width=\linewidth]{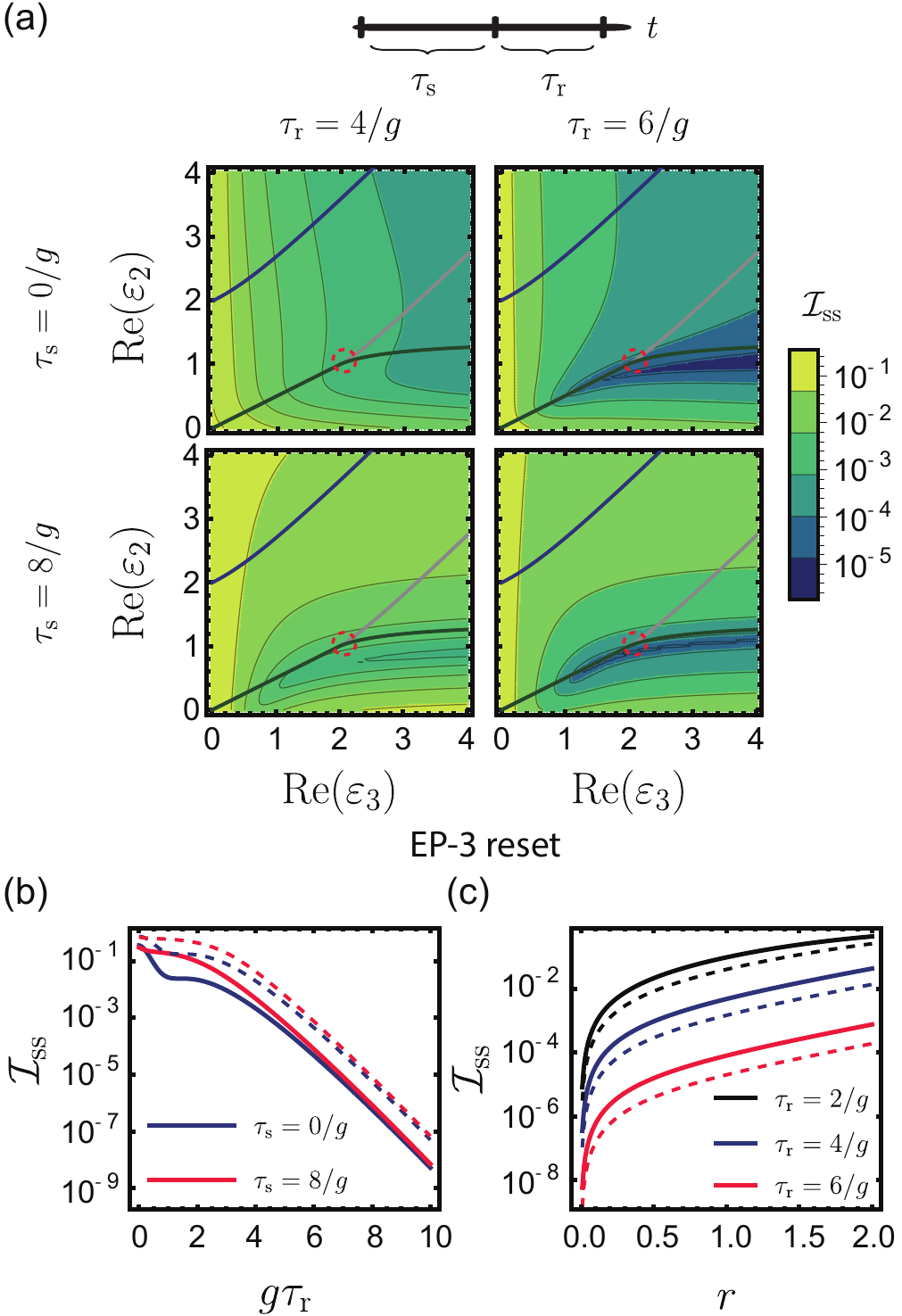}
    \caption{(a) Reset infidelity $\mathcal{I}_{\text{ss}}$ of degenerate resonators as a function of the dimensionless decay rate offsets
    $\Re(\varepsilon_3)$ and $\Re(\varepsilon_2)$ for selected choices of preparation times $\tau_{\text{s}}$ (top and bottom panels) and reset times $\tau_{\text{r}}$ (left and right panels). During the time-interval $\tau_{\text{s}}$, the system is set at the EP-$2$ with $\Re(\varepsilon_3)=0$ and $\Re(\varepsilon_2)=2$. Solid curves on top of the contour plots show the components of the EP branches $++$ (blue), $+-$ (gray), and $--$ (green) in the $\Re(\varepsilon_3)$-$\Re(\varepsilon_2)$ parameter space as in the shaded region of Fig.~\ref{fig:epsa} with EP-$3$ indicated by dashed circles. The other parameters are $\omega_j/g=5000$, $j=1$--$3$, $\kappa_1/g=10^{-3}$, and $r=1$. (b) Reset infidelity $\mathcal{I}_{\text{ss}}$ at the EP-$3$ as a function of reset times $\tau_{\text{r}}$ (in units of $g^{-1}$) for different preparation times $\tau_{\text{s}}$ and decay rate $\kappa_1/g=10^{-3}$. Solid (dashed) curves show data for $r=1.0$ ($r=2.0$). (c) Reset infidelity $\mathcal{I}_{\text{ss}}$ at the EP-$3$ as a function of squeezing parameter $r$ for different reset times $\tau_{\text{r}}$ and for preparation time $\tau_{\text{s}}=8/g$. Solid (dashed) curves show data for $\kappa_1/g=10^{-3}$ ($\kappa_1/g=10^{-1}$). The remaining parameters are chosen as in (a).
    }
    \label{fig:reset}
\end{figure}

Figure~\ref{fig:reset} shows the reset infidelity for different parameter values and for an initial state, which is obtained by waiting for a preparation time $\tau_{\text{s}}$ at EP-$2$ after squeezing the vacuum at resonator $R3$ by a finite $r$. Note that if $\tau_{\text{s}}=0$, one has the initial squeezed state with the covariance matrix given by Eq.~\eqref{eq:V0}, and with $\tau_{\text{s}}=8/g$, one prepares an initial state with entanglement and squeezing split between $R1$ and $R3$, see Fig.~\subref*{fig:dyna}. In Fig.~\subref*{fig:reseta}, we show the dependence of $\mathcal{I}_{\text{ss}}$ on the decay rates $\kappa_2$ and $\kappa_3$ in the region corresponding to the shaded area in Fig.~\subref*{fig:epsa} for the above-mentioned preparation times and immediately following reset times $\tau_{\text{r}}$. Although the regions of low infidelity are relatively broad if all squeezing is concentrated in $R3$ so that no entanglement is present, we observe a narrowing of such regions if $\tau_{\text{s}}=8/g$. These regions tend to cover the EP-$3$ and follow the real components of the $--$ branch of $f(\varepsilon_3)$ as $\varepsilon_3$ is increased. Such a feature is even more prominent for long reset times naturally leading to lower reset infidelities. Note from Fig.~\subref*{fig:epsb} that this branch tends to produce highly dissipative normal modes for $\varepsilon_3>2$. In contrast, at least, one decay rate produced by the $+-$ and $++$ branches is slow even with increasing $\varepsilon_3$, rendering such branches less favorable for the reset. 

Figure~\subref*{fig:resetb} shows the reset infidelity $\mathcal{I}_{\text{ss}}$ as a function of the reset times $\tau_{\text{r}}$ at the EP-$3$ for different initial states. In all displayed cases, low infidelities $\mathcal{I}_{\text{ss}}$ are indeed achieved beyond $\tau_{\text{r}}\sim6/g$, owing to the exponential dependence on $\tau_{\text{r}}$. For such reset times, the distribution of squeezing and entanglement tends to have a minor relative effect on the reset performance. This is in stark contrast with the short-reset-time cases where the decay towards the ground state tends to significantly accelerate if all initial squeezing is poorly distributed, remaining mostly in $R3$. We observe that the reset performance is degraded for small ratios of $\kappa_1/g$ and for increasing initial squeezing parameters as displayed in Fig.~\subref*{fig:resetc}. In such scenarios, for a finite reset time, the infidelity tends to grow asymptotically to unity in the limit $r\to\infty$. 

\section{Discussion}\label{sec:disc}

We observed that fast generation of entanglement and propagation of squeezing in a linear chain of three superconducting resonators may benefit from the detailed understanding of critical damping in the system. Here, the highly dissipative resonator $R2$ acts as an incoherent entanglement generator and squeezing splitter with the cost of reducing the purity of the local states through the increase in their effective temperatures.  
The role of critical damping towards stabilization has also been acknowledged recently in an autonomous quantum thermal machine with two qubits~\cite{Khandelwal2021}.

The stabilization of squeezed states through reservoir engineering in superconducting circuits has been recently reported in ~\cite{Dassonneville2021}. We highlight that the scheme in our paper differs from typical two-mode squeezing operations since it arises from the combination of dissipation and only a single-mode squeezing source available in the beginning of the dynamics, thus, being also distinct from conventional reservoir-engineering protocols. 
On the other hand, we do not need continuous driving terms since the structure of couplings and dissipation of the system promote a separation of time scales for the decay of the normal modes. We explicitly show that this can be beneficial if fine-tuning $\kappa_2$ near a particular EP-$2$ instead of only roughly assuming the conditions $\kappa_j\ll\kappa_2, g$ for $j=1,3$. 

The results shown in Figs.~\ref{fig:dyn} and~\ref{fig:stab} also suggest that concatenating similar structures can be used for fast and stable distribution of entanglement to every other node in a photonic network. Although, spoiling Gaussian features of the system~\cite{Eisert2002,Giedke2002}, entanglement distillation protocols~\cite{Takahashi2010} may be used in such cases to increase the amount of entanglement shared by the nodes. Particular low-order EPs of high-dimensional systems may be used to speed up the generation of quasistable states, and, hence, they may have potential use in the cases in quantum protocols, although the open-system-degeneracy map in such cases becomes more intricate. 

Regarding the unconditional dissipative reset of the system, the role of critical damping becomes more evident. Here, the region near the EP-$3$ and following a particular EP-$2$ branch is a reasonable choice of parameters to produce a substantial performance enhancement of the reset. Since the covariance matrices of the vacuum state and a product of coherent states are identical, such regions in the parameter space could also be used to promote unconditional fast stabilization of coherent states with a proper inclusion of driving terms in the system Hamiltonian.

Let us present typical experimental parameters of the circuit~\subref*{fig:modelb} that could reproduce the findings of this paper. For a resonance frequency of $\omega/(2\pi)=5.0$ GHz, the simulated values of coupling strength and lowest decay frequencies are $g/(2\pi)=1.0$ MHz and $\kappa_1/(2\pi)=1.0$ kHz, respectively. Such resonance frequency and coupling strength have been conveniently experimentally achievable for longer than a decade, and the quality factor of $5\times10^6$ implied by the lowest decay rate can be achieved with state-of-the-art fabrication techniques. The EP-$2$ used for stabilization is, thus, achieved with $\kappa_2/(2\pi)\approx5.66$ MHz and $\kappa_3/(2\pi)=1.0$ kHz, whereas, the EP-$3$ with $\kappa_2/(2\pi)\approx2.83$ and $\kappa_3/(2\pi)\approx5.66$ MHz. Even though the almost four-orders-of-magnitude tunability required to interchange between this particular EP-$2$ and the EP-$3$ may be technically challenging, the maximum achievable decay rates with the QCR are beyond the ones considered here and their demonstrated on/off ratios are close to these requirements~\cite{Moerstedt2022}.

\section{Conclusions}\label{sec:conc}

We demonstrated the theory of exceptional-point-related phenomena for continuous-variable systems described entirely by their second moments, consequently, capturing different nonclassical features and nonlocality largely neglected in previous work. For a linear chain of three lossy superconducting resonators, we analytically obtained its open-system-degeneracy map and observed that different parameter sets yielding different exceptional points can be used to identify sweet spots for the optimization of squeezing propagation, entanglement generation, and reset. 

More precisely, we assessed the role of critical dynamics for dissipative state synthesis by numerically simulating the temporal evolution of the covariance matrix of the system. The region of the parameter space considered in the simulations is physically motivated by recent experimental advances in dissipation-tunable devices embedded to superconducting circuits. 

We found that the quasistabilization into mixed bipartite entangled states generated from an initially squeezed resonator $R3$ is optimized in the vicinities of a particular low-dissipative EP-$2$ produced with symmetric decay rates of resonators $R1$ and $R3$ [see the $++$ branch of $f(\varepsilon)$ in Eq.~\eqref{eq:frealeps}]. In such scenarios, one observed that the time scale for this quasistabilization is minimum for $\kappa_2\approx4\sqrt{2}g$ and $\kappa_1,\kappa_3\ll\kappa_2$. Using the Jordan decomposition of the dynamical matrix, we obtained analytical bounds for the maximum achievable quasistable squeezing-splitting capacity and logarithmic negativity. Remarkably, all residual squeezing of the central resonator is removed within the quasistabilitization timescales, and, consequently, the choice of EP-$2$ also quickly removes the entanglement of $R2$ with the other resonators. 

Furthermore, we investigated the dissipative reset of such nonclassical states to the ground state. The region in the parameter space producing the lowest reset infidelities at given reset times $\tau_{\text{r}}$ requires asymmetric resonator decay rates and tend to follow a particular high-dissipative EP branch, which includes the physically attainable EP-$3$ [see the $--$ branch of $f(\varepsilon)$ in Eq.~\eqref{eq:frealeps}]. In this EP-$3$ case, the distribution of the initial squeezing into the different resonators tends to become irrelevant for the reset performance beyond $\tau_{\text{r}}\sim6/g$. 

In conclusion, this paper paves the way for a deep understanding of the role of exceptional points in multimode continuous-variable systems with potential applications in quantum technology, such as in using dissipation as an ingredient for fast transfer of desired quantum properties. For example, heat engines~\cite{Myers2022} operating with nonequilibrium reservoirs~\cite{Klaers2017} and presenting quantum resources~\cite{Camati2019} arise as systems with promising near-term opportunities.
Moreover, the investigation of exceptional points in such superconducting systems through involved models, see, e.g., Ref.~\cite{Viitanen2021}, is also a potential future line of research. As a final remark, we note that the role of the counter-rotating terms in the system Hamiltonian on the exceptional points may also be addressed with the tools presented in Sec.~\ref{sec:modela}.

\section*{Acknowledgements} \label{sec:acknowledgements}

We acknowledge the Academy of Finland Centre of Excellence Program (Project No.\ 336810), European Research Council under Consolidator Grant No.\ 681311 (QUESS) and Advanced Grant No.\ 101053801 (ConceptQ).

\bibliography{refs}

\appendix
\section{Explicit determination of EPs}{\label{app:EPs}}

Here, we show the characterization of EPs for the open system presented in Sec.~\ref{sec:modelb}, namely, a linear chain of three resonators. Given fixed coupling constants $g$ and parameters of resonator $R1$, we aim at finding the parameters of resonators $R2$ and $R3$ that produce EPs.

We begin by examining the three-mode dynamics in a frame rotating with the angular frequency $\omega_1$ such that $\hrh_{\text{rf}}=\hU\hrh\hU^{\dagger}$, where $\hU=\exp[i\omega_1 t\sum_{j=1}^3\ha_j^{\dagger}{\ha_j}]$. In this frame, the Lindblad master equation~\eqref{eq:LME} describing the studied system becomes
$\textrm{d}\hrh_{\text{rf}}/\textrm{d}t=-i[\hH_{\text{rf}},\hrh_{\text{rf}}]/\hbar+\mathcal{L}_{\downarrow}(\hrh_{\text{rf}})$, where
\begin{equation}
\hH_{\text{rf}}/\hbar=\delta_2\hdgg a_{2}\ha_{2}+\delta_3\hdgg a_{3}\ha_{3}+g(\ha_{1}\hdgg a_{2}+\ha_{2}\hdgg a_{3}+\text{h.c.}),\label{eq:H3rf}
\end{equation}
$\delta_2=\omega_2-\omega_1$, $\delta_3=\omega_3-\omega_1$, and
\begin{align}
    \mathcal{L}_{\downarrow}(\hrh_{\text{rf}})=\frac{1}{2}\sum_{j=1}^{3}\kappa_j\left[2\ha_{j}\hrh_{\text{rf}}\ha_{j}^{\dagger}-\left\{\ha_{j}^{\dagger}\ha_{j},\hrh_{\text{rf}}\right\} \right].\label{eq:LMErf}
\end{align}
Consequently, one can write the dynamical equations for the expectation values $\ev{\ha_j}_{\text{rf}}=\Tr[\hrh_{\text{rf}}\ha_j]$ in a vector notation as $\dot{\mathbf{a}}_{\text{rf}}=-i\mathcal{H}\mathbf{a}_{\text{rf}}$, where $\mathbf{a}_{\text{rf}}=(\ev{\ha_1}_{\text{rf}},\ev{\ha_2}_{\text{rf}},\ev{\ha_3}_{\text{rf}})^{\top}$ and
\begin{align}
    \mathcal{H}=\left(\begin{array}{ccc}
-i\frac{\kappa_{1}}{2} & g & 0\\
g & \delta_2-i\frac{\kappa_{2}}{2} & g\\
0 & g & \delta_3-i\frac{\kappa_{3}}{2}
\end{array}\right),\label{eq:nH3}
\end{align}
as defined in Eq.~\eqref{eq:nH2} of the main text.

To identify the EPs of the three-mode open system, we choose the parametrizations,
\begin{align}
\delta_{2}&=\sqrt{2}g\text{Im}(\varepsilon_{2}),\ \
\delta_{3}=\sqrt{2}g\text{Im}(\varepsilon_{3}),\nonumber\\
\kappa_{2}&=\kappa_{1}+2\sqrt{2}g\text{Re}(\varepsilon_{2}),\ \
\kappa_{3}=\kappa_{1}+2\sqrt{2}g\text{Re}(\varepsilon_{3}),\label{eq:delkap}
\end{align}
so that the effective offsets from $\omega_1$ and $\kappa_1$ are given according to the imaginary and real parts of the complex parameters $\varepsilon_2$ and $\varepsilon_3$.
This allows one to write the characteristic polynomial associated with $\mathcal{H}$, $P(x)=a x^3+b x^2 + c x+d$ with the coefficients,
\begin{align}
    a=&1,\nonumber\\
    b=&\frac{i}{2}[3\kappa_1+2\sqrt{2}g(\varepsilon_2+\varepsilon_3)],\nonumber\\
    c=&-\frac{3\kappa_1^2}{4}-\sqrt{2}g \kappa_1(\varepsilon_2+\varepsilon_3)-2g^2(1+\varepsilon_2 \varepsilon_3),\nonumber\\
    d=&-\frac{i}{8}[\kappa_1^3+8\sqrt{2}g^3\varepsilon_3+2\sqrt{2}g\kappa_1^2(\varepsilon_2+\varepsilon_3)\nonumber\\
    {}&+8 g^2\kappa_1(1+\varepsilon_2\varepsilon_3)].\label{eq:charP}
\end{align}

Given the cubic discriminant,
\begin{align}
    \Delta&=-(4v^3+27w^2),
\end{align}
where
\begin{align}
v&=\frac{3ac-b^2}{3a^2},\ \ w =\frac{2b^3-9abc+27a^2d}{27a^3},\label{eq:disc}
\end{align}
we search for degeneracies in the spectrum of $\mathcal{H}$, which occur when $\Delta=0$. Interestingly, the appearance of EPs depends only on the relationship between $\varepsilon_2$ and $\varepsilon_3$ given by the condition
\begin{align}
    {}&4\varepsilon_2^4 \varepsilon_3^2-8\varepsilon_2^3\varepsilon_3^3+4\varepsilon_2^2(\varepsilon_3^4-5\varepsilon_3^2+1)\nonumber\\
    {}&+4\varepsilon_2(5\varepsilon_3^3-\varepsilon_3)-8\varepsilon_3^4+13\varepsilon_3^2-16=0.\label{eq:EPcond}
\end{align}
Solving Eq.~\eqref{eq:EPcond} for $\varepsilon_2$ yields the four branches,
\begin{align}
    \varepsilon_{2}&=\frac{1}{2}\left[\varepsilon_{3}\pm\sqrt{\frac{\varepsilon_{3}^{4}+10\varepsilon_{3}^{2}-2\pm2\left(1+2\varepsilon_{3}^{2}\right)^{\frac{3}{2}}}{\varepsilon_{3}^{2}}}\right],\label{eq:eps2}
\end{align}
where the signs $\pm$ can be chosen independently.

To identify the order of the EPs, we inspect Eqs.~\eqref{eq:disc} and~\eqref{eq:eps2} more carefully. All EP-$3$'s correspond to the triple root of $P(x)=0$, which is obtained when $v=w=0$. First setting $v=0$ reduces $\varepsilon_2$ in Eq.~\eqref{eq:eps2} to
\begin{align}
    \varepsilon_{2}=\frac{1}{2}\left(\varepsilon_{3}\pm\sqrt{12-3\varepsilon_{3}^{2}}\right),\label{eq:eps2EP3}
\end{align}
and imposing $w=0$ yields $\varepsilon_3\in\{\pm 2,\pm i/\sqrt{2}\}$. Hence, the studied open system presents six distinct EP-$3$, two of which are produced when all resonators are degenerate such that $\varepsilon_3=2\varepsilon_2=\pm2$. The remaining four EP-$3$ are obtained with $\varepsilon_2=(\pm3\sqrt{3}\pm i)/(2\sqrt{2})$ and $\varepsilon_3=2i\textrm{Im}(\varepsilon_2)=\pm i/\sqrt{2}$, thus, requiring frequency shifts from the resonance and equal decay rates for $R1$ and $R3$.

\begin{figure}
    \subfloat{\label{fig:ndepsa}} 
	\subfloat{\label{fig:ndepsb}}
    \centering
    \includegraphics[width=\linewidth]{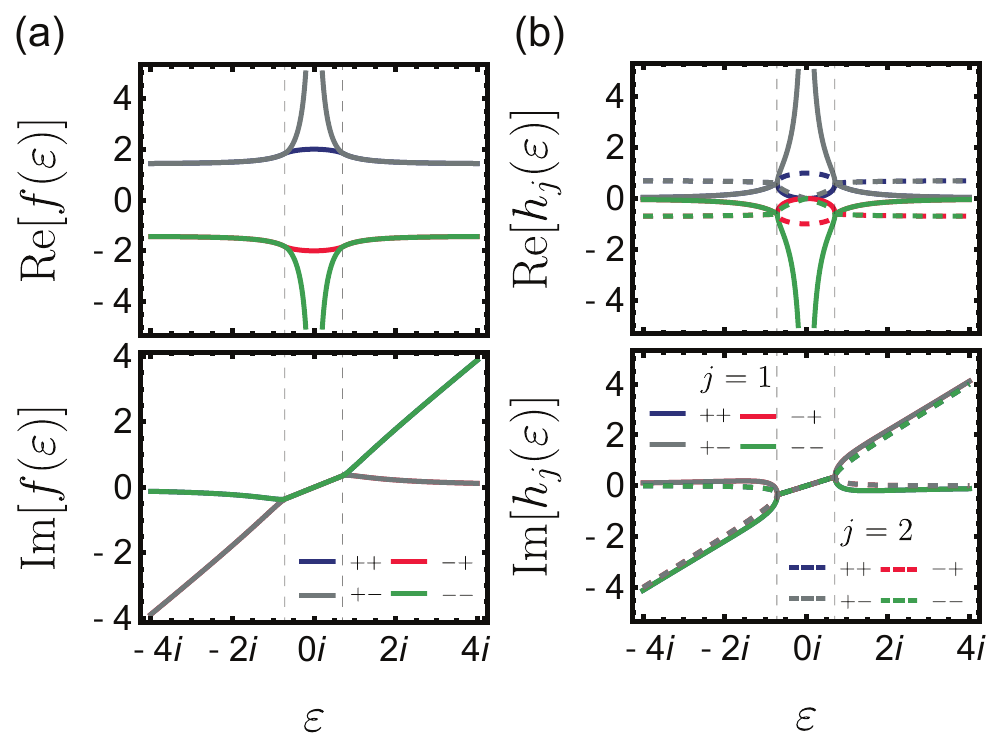}
    \caption{Open-system-degeneracy map for a linear chain of three coupled resonators with degenerate decay rates $\kappa_1=\kappa_3$, which are expressed by a pure imaginary parameter $\varepsilon$. Dependence of (a) $f(\varepsilon)$ and (b) $h_j(\varepsilon)$ on $\varepsilon$. In (b), solid (dashed) curves represent the single (double) root of the characteristic polynomial of $\mathcal{H}$. In all cases, the labels $++$, $-+$, $+-$, and $--$ indicate the four branches of $f(\varepsilon)$ obtained from the corresponding selection of signs in Eq.~\eqref{eq:frealeps}. The vertical dashed lines in all plots highlight the values of $\varepsilon$ yielding EP-$3$.}
    \label{fig:ndeps}
\end{figure}

By defining $\varepsilon=\varepsilon_3$, the parameters of $R2$ produce EPs provided that they are chosen as $\varepsilon_2\equiv f(\varepsilon)$ with $f(\varepsilon)$ defined in Eq.~\eqref{eq:frealeps} of the main text. When degeneracies of $\mathcal{H}$ are present, we express the complex roots of $P(x)=0$ as
\begin{align}
    x_1=\frac{4 a b c-9a^2 d - b^3}{a(b^2-3 a c)},\ \
    x_2=x_3=\frac{9 a d- b c}{2(b^2-3 a c)},
\end{align}
to extract the effective detunings and decay rates of the normal modes as given in Eq.~\eqref{eq:hrealeps}. In Fig.~\ref{fig:ndeps}, we show the rich structure of the branches yielding the EPs for a purely imaginary $\varepsilon$.

Let us comment on the relationship between the spectrum of the non-Hermitian Hamiltonian $\mathcal{H}$ defined in Eq.~\eqref{eq:nH2} and the dynamical matrix $\mathbf{\Gamma}$ defined in Eq.~\eqref{eq:Gamma}. First, we note that $\mathbf{\Gamma}$ also determines the temporal evolution of the mean vector $\mathbf{x}=\ev{\mathbf{\hx}}$ introduced in Sec.~\ref{sec:modela}. We define the $6\times6$ vector of ladder operators $\mathbf{\hA}=[\mathbf{\ha},(\mathbf{\hat{a}^{\dagger}})^{\top}]^{\top}$, where $\mathbf{\ha}=(\ha_1,\ha_2,\ha_3)^{\top}$, $\mathbf{\hat{a}^{\dagger}}=(\hat{a}_1^{\dagger},\hat{a}_2^{\dagger},\hat{a}_3^{\dagger})$. In the Schrödinger picture, the vector of expectation values $\mathbf{A}=\ev{\mathbf{\hA}}$ is obtained from the dynamical equation $\dot{\mathbf{A}}=\text{diag}[-i(\mathcal{H}+\mathcal{H}_1),i(\mathcal{H}^{*}+\mathcal{H}_1^{*})]\mathbf{A}$, where $\mathcal{H}_1=\omega_1\mathbf{I}_3$. By introducing the vector $\mathbf{x}'=(\mathbf{q},\mathbf{p})^{\top}$, where $\mathbf{q}=(\ev{\mathbf{\ha}}+\ev{\mathbf{\ha}}^{*})/\sqrt{2}$ and $\mathbf{p}=-i(\ev{\mathbf{\ha}}-\ev{\mathbf{\ha}}^{*})/\sqrt{2}$, one verifies that $\mathbf{x}'$ is related to $\mathbf{A}$ through a unitary transformation so that $\mathbf{x}'=\mathbf{\Lambda}\mathbf{A}$, where
\begin{align}
    \mathbf{\Lambda}=\frac{1}{\sqrt{2}}\left(\begin{array}{cc}
\mathbf{I}_3 & \mathbf{I}_3 \\
-i\mathbf{I}_3 & i\mathbf{I}_3
\end{array}\right).
\label{eq:Lambda}
\end{align}
Consequently, the dynamical equation for $\mathbf{x}'$ becomes $\dot{\mathbf{x}'}=\mathbf{\Gamma}'\mathbf{x}'$, where
\begin{align}
    \mathbf{\Gamma}'=\mathbf{\Lambda}\left[\begin{array}{cc}
-i(\mathcal{H}+\mathcal{H}_1) & \mathbf{0}_3 \\
\mathbf{0}_3 & i(\mathcal{H}^{*}+\mathcal{H}_1^{*})
\end{array}\right]\mathbf{\Lambda}^{\dagger}.
\label{eq:Gammal}
\end{align}
Since the vectors $\mathbf{x}'$ and $\mathbf{x}$ are equivalent except for the different orderings, the spectrum of $\mathbf{\Gamma}'$ coincides with that of $\mathbf{\Gamma}$, which in the case of open-system degeneracies is given by the eigenvalues defined in Eq.~\eqref{eq:lambdaEP} of the main text.  

\section{Jordan normal form}{\label{app:Jordan}}
In this appendix, we present the explicit Jordan normal form of $\mathbf{\Gamma}$ defined in Eq.~\eqref{eq:Gamma} at some relevant EPs considered in this paper.

\paragraph*{EP-$3$.} We start with the EP--$3$ used for the reset of the system, i.e., the one produced with degenerate resonators ($\omega_1=\omega_2=\omega_3=\omega$), $\kappa_2=\kappa_1+2\sqrt{2}g$, and $\kappa_3=\kappa_1+4\sqrt{2}g$. Using the notation introduced in Sec.~\ref{sec:modela}, the Jordan blocks in the matrix $\mathbf{J} = \text{diag}\left[\mathbf{J}_{\mathbf{s}_1}^{-}(\lambda_{\mathbf{s}_1}^{-}),\mathbf{J}_{\mathbf{s}_1}^{+}(\lambda_{\mathbf{s}_1}^{+})\right]$, and the nonsingular matrix $\mathbf{P}$ read
\begin{align}
    \mathbf{J}_{\mathbf{s}_1}^{\pm}(\lambda_{\mathbf{s}_1}^{\pm})&=\left(\begin{array}{ccc}
\lambda_{\mathbf{s}_1}^{\pm} & 1 & 0\\
0 & \lambda_{\mathbf{s}_1}^{\pm} & 1\\
0 & 0 & \lambda_{\mathbf{s}_1}^{\pm}
\end{array}\right),\ \ \mathbf{s}_1=(1,3),
\label{eq:JB3}
\end{align}
\begin{align}
\mathbf{P}=\left(\begin{array}{cccccc}
-i & -\frac{i\sqrt{2}}{g} & -\frac{i}{g^{2}} & i & \frac{i\sqrt{2}}{g} & \frac{i}{g^{2}}\\
-1 & -\frac{\sqrt{2}}{g} & -\frac{1}{g^{2}} & -1 & -\frac{\sqrt{2}}{g} & -\frac{1}{g^{2}}\\
-\sqrt{2} & -\frac{1}{g} & 0 & -\sqrt{2} & -\frac{1}{g} & 0\\
i\sqrt{2} & \frac{i}{g} & 0 & -i\sqrt{2} & -\frac{i}{g} & 0\\
i & 0 & 0 & -i & 0 & 0\\
1 & 0 & 0 & 1 & 0 & 0
\end{array}\right),\label{eq:PEP3}
\end{align}
where $\lambda_{\mathbf{s}_1}^{\pm}$ are given as in Eq.~\eqref{eq:lambdaEP} of the main text with $\varepsilon=2$ and $f(\varepsilon)=1$. Consequently, one obtains
\begin{align}
\text{e}^{\mathbf{J}t}&=\bigoplus_{m=\mp}\text{e}^{\lambda_{\mathbf{s}_1}^{m}t}\left(\begin{array}{ccc}
1 & t & t^{2}/2\\
0 & 1 & t\\
0 & 0 & 1
\end{array}\right).\label{eq:eJtEP3}
\end{align}
\paragraph*{EP-$2$.}  Let us consider the EP-$2$ used for the quasistabilization of squeezing and entanglement, i.e., the one obtained with degenerate resonators, $\kappa_2=\kappa_1+4\sqrt{2}g$, and $\kappa_3=\kappa_1$. The Jordan blocks here defining the matrix $\mathbf{J} = \text{diag}\left[\mathbf{J}_{\mathbf{s}_1}^{-}(\lambda_{\mathbf{s}_1}^{-}),\mathbf{J}_{\mathbf{s}_2}^{-}(\lambda_{\mathbf{s}_2}^{-}),\mathbf{J}_{\mathbf{s}_1}^{+}(\lambda_{\mathbf{s}_1}^{+}),\mathbf{J}_{\mathbf{s}_2}^{+}(\lambda_{\mathbf{s}_2}^{+})\right]$ are \begin{align}
    \mathbf{J}_{\mathbf{s}_1}^{\pm}(\lambda_{\mathbf{s}_1}^{\pm})&=\lambda_{\mathbf{s}_1}^{\pm},\ \ \mathbf{s}_1=(1,1),\nonumber \\
\mathbf{J}_{\mathbf{s}_2}^{\pm}(\lambda_{\mathbf{s}_2}^{\pm})&=\left(\begin{array}{cc}
\lambda_{\mathbf{s}_2}^{\pm} & 1\\
0 & \lambda_{\mathbf{s}_2}^{\pm}\end{array}\right),\ \ \mathbf{s}_2=(2,2),
\label{eq:JB2}
\end{align}
and the singular matrix $\mathbf{P}$ reads
\begin{align}
\mathbf{P}=\left(\begin{array}{cccccc}
-i & i & 0 & i & -i & 0\\
-1 & 1 & 0 & -1 & 1 & 0\\
0 & \sqrt{2} & -\frac{1}{g} & 0 & \sqrt{2} & -\frac{1}{g}\\
0 & -i\sqrt{2} & \frac{i}{g} & 0 & i\sqrt{2} & -\frac{i}{g}\\
i & i & 0 & -i & -i & 0\\
1 & 1 & 0 & 1 & 1 & 0
\end{array}\right),\label{eq:PEP2}
\end{align}
where $\lambda_{\mathbf{s}_j}^{\pm}$ are given as in Eq.~\eqref{eq:lambdaEP} of the main text with $\varepsilon=0$ and $f(\varepsilon)=2$. In this case, the Jordan decomposition of $\mathbf{\Gamma}$ gives rise to
\begin{align}
\text{e}^{\mathbf{J}t}&=\bigoplus_{m=\mp}\left(\begin{array}{ccc}
\text{e}^{\lambda_{\mathbf{s}_1}^{m}t} & 0 & 0\\
0 & \text{e}^{\lambda_{\mathbf{s}_2}^{m}t} & \text{e}^{\lambda_{\mathbf{s}_2}^{m}t}t\\
0 & 0 & \text{e}^{\lambda_{\mathbf{s}_2}^{m}t}
\end{array}\right).\label{eq:eJtEP2}
\end{align}
This decomposition is employed in Eq.~\eqref{eq:VsolJ} to analytically obtain the time-evolved covariance matrix $\mathbf{V}$ and, consequently, the theoretical bounds presented in Sec.~\ref{sec:resultsa}.



\end{document}